\documentclass[a4paper,10pt]{svmult}
\usepackage[utf8x]{inputenc}
\usepackage{graphicx}

\usepackage{multicol}        % used for the two-column index
\usepackage[bottom]{footmisc}% places footnotes at page bottom

\begin{document}
\title*{
 Electric Field Effects on Graphene Materials}

\titlerunning{Electric Field Effects on Graphene Materials}

\author{
Elton J. G. Santos}

\institute{
Elton  J. G. Santos$^{\dagger,\star}$ \at
School of Engineering and 
Applied Sciences, Harvard University, Cambridge, Massachusetts 02138, USA.\\
$^{\dagger}$Present address: Department of Chemical Engineering, Stanford University, Stanford, California 94305, USA. \\ 
$^\star$\email{eltonjos@stanford.edu} \\
}

\maketitle

\abstract{
Understanding the effect of electric fields on the physical and chemical properties of 
two-dimensional (2D) nanostructures is instrumental 
in the design of novel electronic and optoelectronic devices. 
Several of those properties are characterized in terms of the dielectric constant which 
play an important role on capacitance, conductivity, screening, dielectric losses and refractive index. 
Here we review our recent theoretical studies using density functional 
calculations including van der Waals interactions 
on two types of layered materials of similar two-dimensional 
molecular geometry but remarkably 
different electronic structures, that is, graphene and molybdenum disulphide (MoS$_2$).  
We focus on such two-dimensional crystals because of they complementary physical and chemical 
properties, and the appealing interest to incorporate them in the next generation of electronic and 
optoelectronic devices. We predict that the effective dielectric constant ($\varepsilon$) 
of few-layer graphene and MoS$_2$ is tunable by external electric fields ($E_{\rm ext}$). 
We show that at low fields ($E_{\rm ext}^{}<0.01$ V/\AA)  
$\varepsilon$ assumes a nearly constant value $\sim$4 for both materials, but increases at 
higher fields to values that depend on the layer thickness. 
The thicker the structure 
the stronger is the modulation of $\varepsilon$ with the electric field. 
Increasing of the external field perpendicular to the layer surface 
above a critical value can drive the systems to an unstable state where the 
layers are weakly coupled and can be easily separated. 
The observed dependence of $\varepsilon$ on the external field 
is due to charge polarization driven by the bias, which show several 
similar characteristics despite of the layer considered. 
All these results provide key information about control and understanding of the screening 
properties in two-dimensional crystals beyond graphene and MoS$_2$. 
}

%------------------------------------------------------------------------------
\section{Introduction}
\label{intro}
%------------------------------------------------------------------------------

%
%%

Electron-electron interactions play a central role on a wide range of 
electronic phenomena in two-dimensional (2D) materials. 
One of the main ingredients that determines the Coulomb interaction strength in 
those systems is screening, which can be characterized by dielectric constant, $\varepsilon$. 
Indeed, screening effects based on $\varepsilon$ play a fundamental role in determining the 
electron dynamics, the optical exciton binding energy, the electron and hole mobilities 
as well as charge storage features. In this context the Coulomb interactions are confined 
in a two-dimensional geometry which can give place to a new set of dielectric 
properties depending on the electronic nature of the 2D crystal. 

Graphene, a semimetal with zero bandgap, and 
MoS$_2$, a low-dimension direct band gap semiconductor, are two 
representative members of the 2D family that have been receiving much attention 
in many fields due to their remarkable chemical 
and physical properties (Castro Neto et al. 2009, Wang et al. 2012). 
One of the main features that influences all these properties is the 
layer thickness, which determines the charge distribution in the device 
as well as the electronic structure through the band gap. 
In particular, graphene can become a semiconductor with a band gap of several tenths of meV's 
at a bilayer structure subjected to high electric gate bias (Castro et al. 2007). 
MoS$_2$, in its turn, has a sizable band gap that varies as a function 
of the number of layers which reaches values of about 1.8 eV at the monolayer limit (Mak et al. 2010). 
In both situations the electric-field screening is observed to 
change as the dielectric response depends on the intrinsic electronic 
properties as well as on $\varepsilon$.

In fact, the large range of values for $\varepsilon$ found by 
different experiments on 
graphene (Elias et al. 2011, Siegel et al. 2011, Bostwick et al. 2010, Reed at al. 2010, Wang et al. 2012, Sanchez-Yamagishi et al. 2012, Fallahazad et al. 2012, Jellison et al. 2007)
and MoS$_{2}$ (Zhang et al. 2012, Kim et al. 2012, Bell et al. 1976, Frindt et al. 1963, Beal et al. 1979)
has become a subject of considerable discussions. 
More factors, apart from the external electric 
fields and layer thickness, indicate that $\varepsilon$ might  
depend on the underneath substrate as recently measured for graphene (Hwang et al. 2012)
and MoS$_2$ layers (Bao et al. 2013). 
In practical terms, the dielectric
constant is defined by $\varepsilon=(\varepsilon_{sub} + \varepsilon_{vac})$/2, 
with $\varepsilon_{sub}$ and $\varepsilon_{vac}$ the dielectric 
constant values for the substrate and vacuum, respectively. However, 
this approach suggests that the environment could play a role in the determination 
of the intrinsic dielectric constant of these 2D materials. Therefore, 
it is paramount to determine the intrinsic value of $\varepsilon$ despite 
of external screening environments. 

In the present Chapter, we provide a review of some of our
recent computational studies on the effect of electric fields on multilayer 
graphene and MoS$_2$. We will consider both layered systems at different field magnitudes (0-1.0 V/\AA) 
and number of layers (2L-10L). Some differences on the dielectric response
between graphene and MoS$_2$ will be discussed based on simple electrostatic concepts, which will 
give generality to the calculations for other 2D-layers still to be explored.

\section{Electrical Field Tuning of the Dielectric Constant}
\label{dielectric-constant}

\begin{figure}
\includegraphics[width=4.300in]{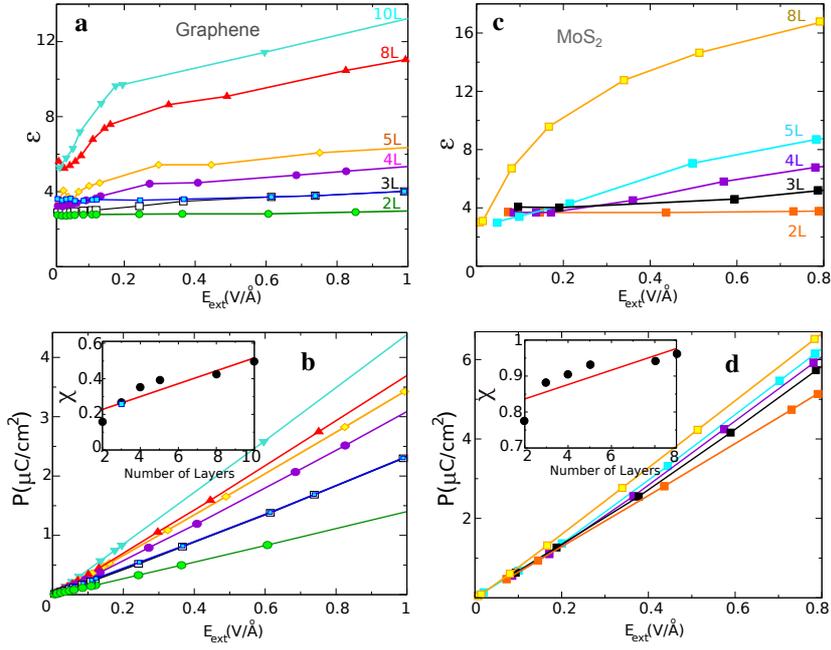}%\centering
\caption{
Calculated $\varepsilon$ and electric polarization P($\mu$C/cm$^2$) 
as a function of E$_{\rm ext}$ for graphene and MoS$_2$ structures. 
Results for graphene, {\bf a} and {\bf b}, and for MoS$_2$, {\bf c} and {\bf d}, 
are show in the range of 2L$-$10L. The insets in {\bf b}, and {\bf d}, show 
the electric susceptibility $\chi$ versus the number of layers. 
A linear fitting is shown by the red line. 
Calculations for graphene were performed using the Bernal stacking order, 
which has not shown appreciable changes as compared 
to the Rhombohedral stacking. 3L is also shown in Rhombohedral stacking using filled 
blue squares in {\bf a}. The Bernal stacking was used for all calculations on MoS$_2$. 
Adapted from (Santos et al. 2013a, Santos et al. 2013b). 
}
\label{fig1}
\end{figure}

Figure \ref{fig1}{\bf a}, {\bf c} display how $\varepsilon$ evolves with external 
fields for different number of graphene and MoS$_2$ layers, respectively. 
At low fields, $E_{\rm ext}\leq$0.001 V/\AA, $\varepsilon$ is almost 
independent of the number of layers having a value close 
to $\sim 4$ for both two-dimensional crystals.  
As the external field $E_{\rm ext}$ is increased, 
$\varepsilon$ reaches larger values, up to $\varepsilon=$12.0 
at E$_{\rm ext}=$1.0 V/\AA~for $N=10$ graphene layers. 
Similar electric response is observed 
for MoS$_2$ with $\varepsilon=$16.8 at $E_{\rm ext}=$0.8 V/\AA~for 
$N=8$ layers with an approximately linear dependence of $\varepsilon$ 
on the number of layers at a fixed magnitude of the field. 
These values for $\varepsilon$ are in good agreement with those found 
by experimental groups working on graphene (Elias et al. 2011, Siegel et al. 2011, Bostwick et al. 2010, Reed at al. 2010, Wang et al. 2012, Sanchez-Yamagishi et al. 2012, Fallahazad et al. 2012, Jellison et al. 2007)
and on MoS$_2$(Zhang et al. 2012, Kim et al. 2012, Bell et al. 1976, Frindt et al. 1963, Beal et al. 1979) samples. 
The electric susceptibility $\chi^{}$ extracted from the polarization $P$  
clearly shows the roughly linear dependence on the number of layers $N$ as plotted 
in Fig. \ref{fig1}{\bf b}, {\bf d}. Moreover, the 
comparison between graphene and MoS$_2$ also gives that the 
tuning of $\varepsilon$ with the external field is 
larger to the latter. At $N=8$, $\chi^{}$ is $\chi=$0.95 for multilayer MoS$_2$ and 
$\chi=$0.42 for multilayer graphene, which is less than half of that value 
calculated for MoS$2$. This suggests that the dichalcogenide layer is more electrically 
polarizable than graphene. 

We note that electric fields of the magnitude considered here
can in principle be experimentally created, 
as recently obtained in the case of 3L graphene (Zou et al. 2013) which fields 
close to $\sim$0.6 V/\AA were achieved taking into account HfO$_2$ gates. In the case of 
MoS$_2$, the high dielectric breakdown, due to the chemical character of the Mo$-$S covalent bonds, 
allows the application of large electric bias as recently
reported in voltage-current measurements (Lembke et al. 2012).

\section{Interlayer Electric Field: Spatial Dependence}
\label{effective-efiled}

\begin{figure}
\includegraphics[width=4.7in]{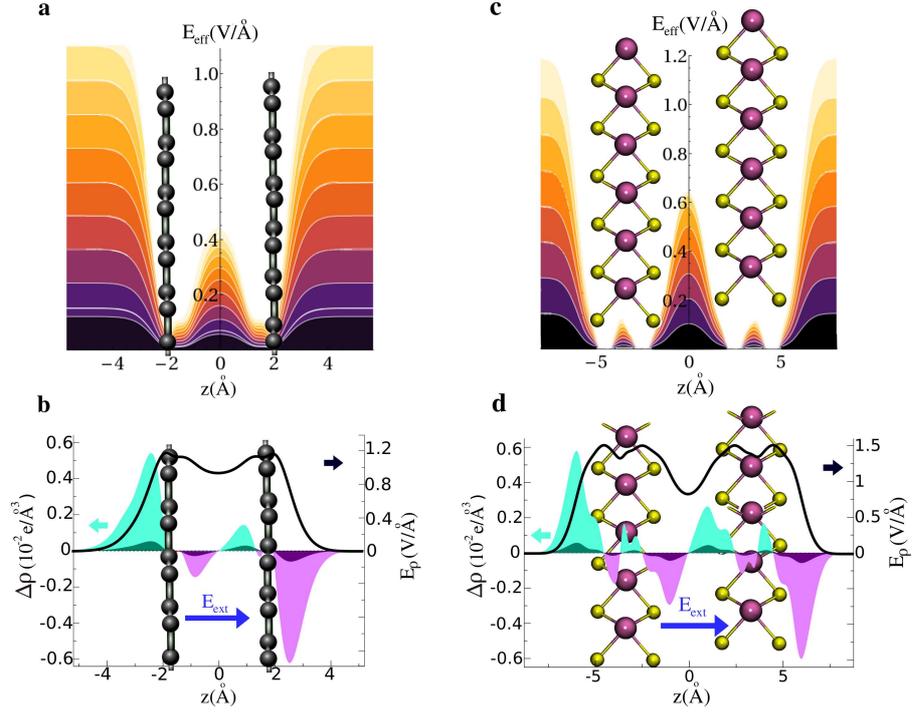}%\centering
\caption{Effective electric field $E_{\rm eff}^{}$ calculated 
as a function of the interlayer distance at different external 
fields $E_{\rm ext}^{}$ for {\bf a}, bilayer graphene, 
and {\bf c}, bilayer MoS$_2$. The color gradient shows the evolution from low (dark color) 
to high (bright color) values of the applied electric fields. 
Induced charge densities, 
$\Delta\rho^{}=\rho({\rm E}_{\rm ext}) - \rho(0)$, in e/\AA$^{3}$, 
between the {\bf b}, carbon surfaces and the {\bf d}, MoS$_2$-planes. 
For graphene, 
the bolder and lighter shaded curves correspond to E$_{\rm ext}=$ 0.12 V/\AA~and 
E$_{\rm ext}=$ 1.22 V/\AA, respectively. 
The solid black curve corresponds to the electric field generated 
by the induced charge, E$_{\rho}$, at E$_{\rm ext}=$ 1.22 V/\AA.
For MoS$_2$, the bolder and lighter shaded curves correspond to $E_{\rm ext}=$ 0.14 V/\AA~and 
$E_{\rm ext}=$ 1.5 V/\AA, respectively.
The solid black curve corresponds to $E_{\rho}^{}$ 
at $E_{\rm ext}=$ 1.5 V/\AA.
The large blue arrow shows the direction of $E_{\rm ext}^{}$ relative to the bilayer structures. 
Adapted from (Santos et al. 2013a, Santos et al. 2013b). 
}
\label{fig2}
\end{figure}

Next we discuss the origin 
of the electric-field mediated tunable dielectric constant 
in graphene and MoS$_2$ layered systems. 
Figure \ref{fig2} shows the electric response in terms of the effective electric field $E_{\rm eff}^{}$
calculated from the Hartree potential $V_H$ along the supercell for bilayer structures.
The application of the external field $E_{\rm ext}^{}$ generates an interlayer charge-transfer which 
partially cancels $E_{\rm ext}^{}$ inducing the appearance of $E_{\rm eff}^{}$ 
in the region between the layers. At low $E_{\rm ext}$, all the induced values of $E_{\rm eff}$ 
are approximately constant, within the numerical accuracy of our model, assuming 
similar shapes as displayed in the dark regions of Fig. \ref{fig2}{\bf a},{\bf c}. 
At fields close to those used to modify the band gap 
of 2L graphene (Mak et al. 2009, Zhang et al. 2009, Castro et al. 2007), 
or used in MoS$_2$ transistors (Radisavljevic et al. 2011), that is 
$E_{\rm ext}=$0.08 V/\AA, the effective field E$_{\rm eff}$ is already  
dependent on position z, with a maximum at the mid-point between the layers. MoS$_2$ has the 
difference to be formed by S$-$Mo$-$S bonds perpendicular to the external field, which induce 
a smaller but finite contribution between the S atoms. Moreover, the effective field on 2L 
MoS$_2$ assumes a narrower a shape relative to graphene with negligible values close to S. 
The electric response can also be analyzed based on the 
the induced charge densities, $\Delta\rho$, 
at different fields as plotted in Figure \ref{fig2}{\bf b},{\bf d}. Both layered systems 
show a charge accumulation at the layer that is under positive potential $+V$
and a corresponding depletion at the other one $-V$. 
The integration of $\Delta\rho$ along z, utilizing the Poisson equation  
$\nabla^{2}V(z)=-\Delta \rho/\varepsilon_{0}$, where 
$\varepsilon_{0}$ is the vacuum permittivity,  
results in a response electric field $E_{\rho}^{}$ (solid black 
line in  Figure \ref{fig2}{\bf b},{\bf d}) 
that screens the external electric field, that is, 
$E_{\rm eff}\approx$ $E_{\rm ext}$ - $E_{\rho}^{}$.

\section{Electric Field Damping in Multilayer Systems}
\label{Damping-multilayer}

\begin{figure}
\includegraphics[width=4.500in]{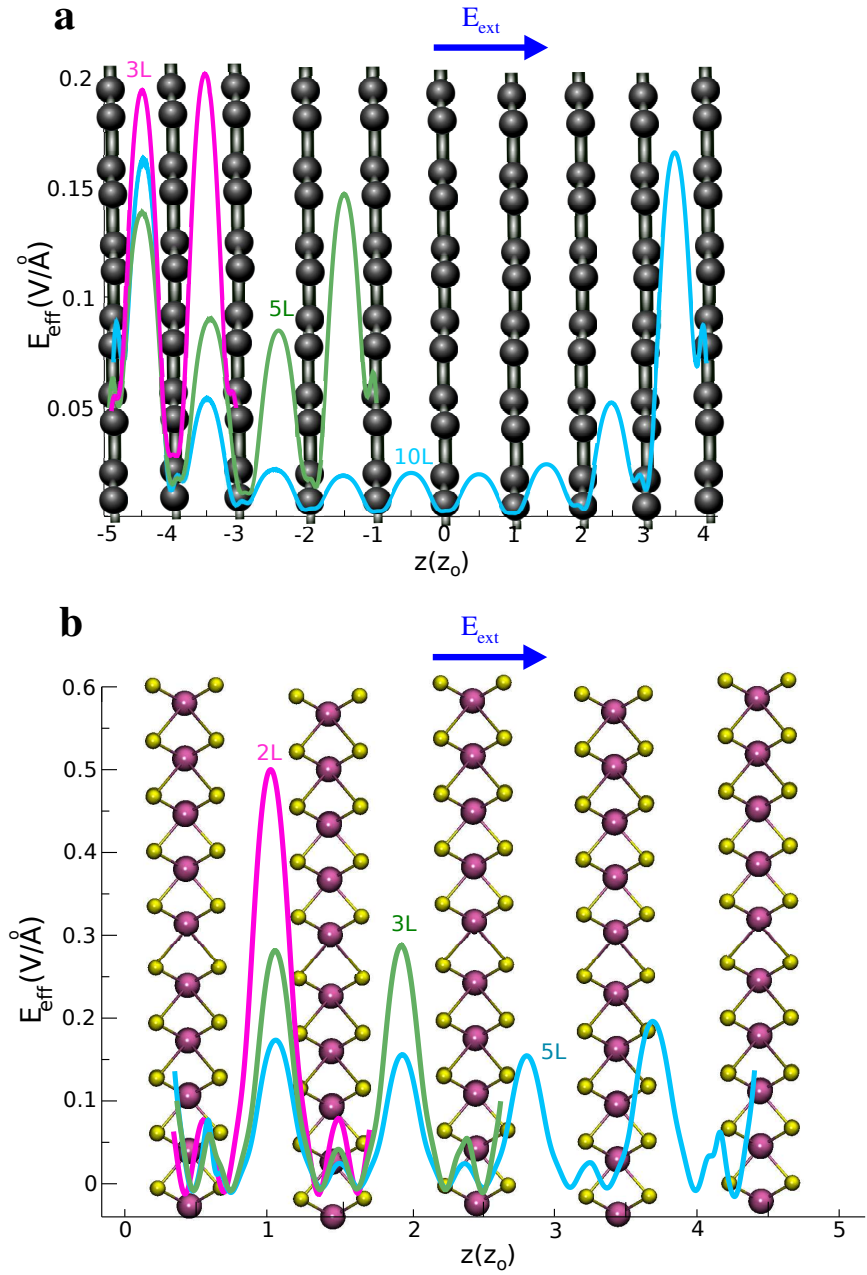}\centering
\caption{Effective electric field $E_{\rm eff}$ as a function of the interlayer position 
$z_{\rm o}$ for {\bf a}, 3L-10L graphene and {\bf b}, 2L-5L MoS$_2$. Note that $z_{\rm o}$ assumes 
different values for graphene ($z_{\rm o}=$3.41 \AA) and MoS$_2$ ($z_{\rm o}=$6.70 \AA). Geometries are 
shown on the background on each panel. The applied fields are 
0.50 V/\AA~and 0.73~V/\AA~for graphene and MoS$_2$, respectively. 
The blue arrow indicates the relative orientation of $E_{\rm ext}$. 
Adapted from (Santos et al. 2013a, Santos et al. 2013b). 
}
\label{fig3}
\end{figure}
In the previous section, 
we have considered in detail the 
the electric response due to external fields on graphene and MoS$_2$ 2L structures. 
Although this is an important system, other aspects are also crucial to understand and control 
the screening associated to two-dimensional crystals. For example, one needs to explore 
the characteristics of multilayer systems subjected to external bias, as well as the possibility
to compare structures with different electronic character. 
This kind of knowledge is instrumental in possible applications in electronics and optoelectronics. 

We address next the dependence of $\varepsilon_{}$ as a function of the number of graphene and MoS$_2$ 
layers $N$ as shown in Figure \ref{fig3}.  
Despite the electronic character of each system, the application of E$_{\rm ext}$ on thicker 
structures creates higher E$_{\rm eff}$ in the first few layers with 
a reduction of field in the innermost regions of the structure. 
For example in graphene, in the $N=10$ case, the maximum value of  
E$_{\rm eff}$ between the two carbon layers at $z=3z_{z_o}$ and $4z_{o}$ is 3.2 times larger 
than that between the layers at $z=2z_{o}$ and $3z_{o}$. In deeper layers, the field decays further 
reaching even smaller values. For MoS$_2$, the outermost layers show a slightly higher E$_{\rm eff}$ and the field 
decay follows that observed for graphene. However, the damping between internal and external layers, 
that is, those closer to the gate bias, is slightly different as that observed at the carbon planes. 
This suggests that the charge polarization of the layers due to the 
different electronic character (Fig. \ref{fig1}{\bf c},{\bf d}) plays 
an important role on the screening behavior and also on the electrical 
tuning of the dielectric constant with the number of layers.   
As $\varepsilon_{}$ is calculated by the ratio of the external and internal fields to the slabs, 
the enhancement in the value of the dielectric constant with the number of layers $N$
is directly related to the reduction of the field in the 
innermost regions of the structure which leads to lower 
$\varepsilon$ values for lower values of $N$. 
This decay of field with the layer thickness is in 
good agreement with recent electrostatic force microscopy and 
Kelvin probe microscope measurements performed 
for MoS$_2$ (Castellanos-Gomez et al. 2013, Li et al. 2013) and graphene (Datta et al. 2009).

\section{Electrostatic Exfoliation on Graphene and MoS$_2$ Layers}
\label{exfoliation}

\begin{figure}
\includegraphics[width=4.700in]{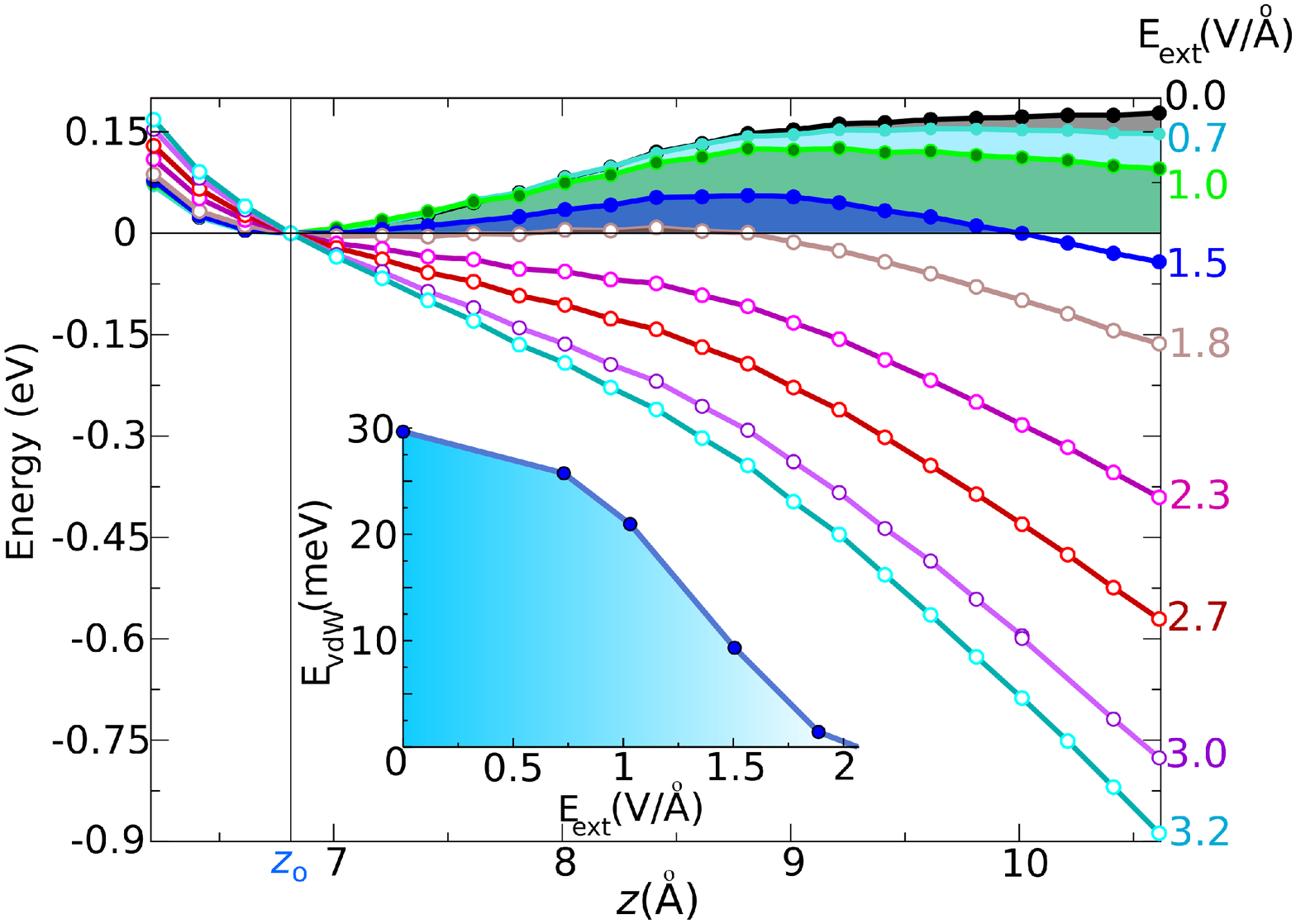}\centering
\caption{Energy per MoS$_2$ unit cell as a 
function of interlayer distance for different values of $E_{\rm ext}$ (V/\AA). 
The vertical solid line indicates the equilibrium interlayer distance 
{\it z}$_{o}=$6.70 \AA. The inset 
shows the van der Waals barrier ($E_{\rm vdW}$) per atom as 
a function of $E_{\rm ext}^{}$. Adapted from (Santos et al. 2013b).}
\label{fig4}
\end{figure}

In this section we analyze the possibility to use an electrostatic gate 
that can be used to exfoliate graphene and MoS$_2$ layers at different bias. 
We note that there is 
an upper limit on the magnitude of $E_{\rm ext}$ that can be applied to the systems as the bias 
induces a shift of the 
equilibrium position of the layers to higher interlayer separations.  
Figure \ref{fig4}~displays the total energy for MoS$_2$ 2L as a function of the interlayer 
distance $z$. We focus on MoS$_2$ since similar effects are observed for graphene. 
At $E_{\rm ext}^{}=$0.0 V/\AA, a van der Waals barrier $E_{\rm vdW}$ of 
30 meV/atom prevents the separation of the two 
layers from {\it z}$_{o}$ to infinity. At finite $E_{\rm ext}$, the value of $E_{\rm vdW}$ 
decreases, indicating that the MoS$_2$-layers become less bound.  
At $E_{\rm ext}=$ 2.0 V/\AA, 
the two dichalcogenide layers can be easily separated 
with a barrier of only 0.45 meV/atom. 
This indicates that an electrostatic gate can be utilized for exfoliating and printing 
few-layers MoS$_2$ in pre-pattern form similarly to that  
observed for graphene (Lian et al. 2009). Since several challenges of making industrially 
available large-scale areas of 2D-crystals and fabricating atomic features with precise 
electronic structure are still to be overcome, the electrostatic exfoliation shown in our calculations 
could open new avenues for the achievement of such desired properties.

\section{Conclusions}
\label{conclusions}

In this Chapter we have reviewed the electrical response of two representative layered materials 
for future devices-based on graphene and MoS$_2$. We have focused on the interplay 
between electric fields and screening properties of few-layer structures. Density functional 
theory was the main tool used to compute the properties of the analyzed systems. 
We have used simple models to understand the observed trends. In particular, 
we find that the effective dielectric constant of graphene and MoS$_2$
is electrically tunable, with the layer thickness playing an important role in the enhancement of
the effect. The thicker the structure is, the stronger the modulation with electric fields. 
The driving force for such behavior is due to the linear dependence of the electrical polarization 
of the layers on the external field. The response field computed from the polarization charge 
does not screen completely the external bias, which generate higher interlayer 
fields at thinner structures. Differences due to semi-metallic and semiconducting electronic character
of the layers are observed in terms of the field damping inside of the compounds: graphene tends 
to screen the external field at the outermost layers of system, while MoS$_2$ the field penetrates 
deeper in the layers. These results are in sound agreement with recent experiments performed for both materials.  

We have also explored the possibility to control the layer exfoliation using electric fields. 
We have found that the induced interlayer charge imbalance generated by the bias can drive 
the system to an unstable state where the layers can be separated from each other. 
The interlayer equilibrium position is modified as a function of the field magnitude, which induces 
a reduction of the van der Waals barrier that keeps the layers together. As a result, there are 
variations of the interlayer separations even at low-fields. This investigation is highly relevant 
in the interpretation of experimental results underway since the field of 2D-materials is just in 
its beginning where several techniques and effects are still to 
be developed and explored.

\end{document}